\documentclass[12pt,a4paper]{article}
\usepackage[latin1] {inputenc}
\usepackage[T1]{fontenc}
\usepackage{amsmath}
\usepackage{amsfonts}
\usepackage{amssymb}
\usepackage[english]{babel}
\usepackage[dvips]{graphics,color}
\usepackage{relsize}
\usepackage{graphicx,psfrag,amsmath,amssymb,amsfonts,bbm,latexsym,color,dcolumn,
epsf}

\begin{document}
\title{Non-superposition effects in the Dirichlet Casimir effect}
\author{C.~Ccapa Ttira, C.~D.~Fosco and E.~L.~Losada
\\
{\normalsize\it Centro At\'omico Bariloche and Instituto Balseiro}\\
{\normalsize\it Comisi\'on Nacional de Energ\'\i a At\'omica}\\
{\normalsize\it R8402AGP Bariloche, Argentina.}\\
}

\date{}
\maketitle
\begin{abstract}
\noindent 
We study non-superposition effects in the Dirichlet Casimir interaction
energy for $N$ boundaries in $d$ spatial dimensions, quantifying its
departure from the case of an interaction where a superposition principle
is valid.  We first derive some general results about those effects, and
then show that they only become negligible when the distances between
surfaces are larger than the sizes of each individual surface.  We consider
different examples in one, two and three spatial dimensions.
\end{abstract}
\section{Introduction}\label{sec:intro}
Among the many interesting and distinctive features of the Casimir
effect~\cite{rev}, not the least important is the fact that the forces it
produces between (more than two) conducting surfaces do not satisfy, in
general, a superposition principle. In other words, when dealing with more
than two objects, the interaction energy cannot be written as the sum of
the interaction energies corresponding to all the possible object pairs.
As a consequence, knowledge of the energy of a system before the addition
of an extra surface may seem to be of little help, since there is no
obvious way to include the extra interaction terms.  Not unexpectedly, this
non-superposition property is shared by the van der Waals
interaction~\cite{milonni}. 

Quite apart from its theoretical interest,
it should be noted that this kind of phenomenon may also be of practical
relevance, since it could be helpful in some approximate calculation
schemes, in situations where nonlinear effects are small.  For example, if
there exists a regime where superposition  is approximately valid, one
should expect the dominant term in the Casimir energy to be akin to a two
body interaction potential, albeit with a non-Coulombian potential, plus
corrections. Under some assumptions, those corrections can be, as we shall
see, small perturbations.

In this paper, we first investigate the non-superposition effects in a quite
general approach. We then argue that when those effects are small, a
perturbative expansion naturally suggests itself.  We then discuss and
apply that approximation within the context of  different examples.

This paper is organized as follows: in section~\ref{sec:method} we first
review the functional approach to the calculation of the Casimir energy.
In~\ref{sec:nonsup}, we deal with the study of non-superposition effects,
relating them to a perturbative expansion in~\ref{sec:perturbative},with
examples in $1+1$,  $2+1$, and $3+1$ dimensions.  

\noindent In~\ref{sec:concl}, we present our conclusions.
\section{The method}\label{sec:method}
In order to analyze properties of the Casimir energy, it is convenient
to introduce one of its concrete representations. We shall use here one
that is based on the functional integral formalism introduced 
in~\cite{kardar,kardar2,recent}.

\noindent What follows is a review of its main aspects, adapted to the problem at hand.

Denoting by ${\mathcal Z}\big[\{ \Sigma^{(a)}\}\big]$ the Euclidean vacuum
amplitude for a real massless scalar field $\varphi$ in the
presence of $N$ Dirichlet surfaces $\Sigma^{(a)}$, $a=1,\ldots,\, N$, the
total vacuum energy $E_0$ may be written as follows:
\begin{equation}\label{eq:e0}
E_0 \;=\; - \, \lim_{T \to \infty} \left\{
\frac{1}{T} \,\log \frac{{\mathcal Z}\big[\{ \Sigma^{(a)}\}\big]}{{\mathcal Z}_0}
\right\} \;,
\end{equation}
where $T$ is the extent of the (imaginary) time interval, and ${\mathcal
Z}_0$ is the free (no surfaces) vacuum amplitude. The role of the latter
is just to fix the vacuum energy to zero when there are no surfaces.

On the other hand, the vacuum amplitude may be written as a
functional integral\footnote{In our use of the functional integral
formalism, we follow the approach and conventions of~\cite{zinn}.}: 
\begin{equation}\label{eq:defzsiga} 
{\mathcal Z}\big[\{\Sigma^{(a)}\}\big] \,=\, \int [{\mathcal D}\varphi]\;
e^{-S_0(\varphi)}\;,
\end{equation}
where $S_0$ is the free Euclidean action, which for a massless scalar field
in $d+1$ dimensions reads: $S_0(\varphi)=\frac{1}{2} \int
d^{d+1}x\;(\partial\varphi)^2$, and $[\mathcal{D}\varphi]$ denotes the path
integral measure corresponding to a scalar field which satisfies Dirichlet
boundary conditions on each surface $\Sigma^{(a)}$. 

It is quite useful to write that measure in the equivalent way:
\begin{equation}\label{eq:defzsiga1}
[{\mathcal D}\varphi]\;=\; {\mathcal D}\varphi \;\times\; \prod_{a=1}^N
\,\delta_{\Sigma^{(a)}}\big[\varphi\big] \;,
\end{equation}
where we introduced a $\delta$-functional for the field on each
surface. 

In what follows, we focus on the $d=3$ case, although every step will have
its analogue for different numbers of dimensions. The changes to the final
expressions required to deal with $d\neq3$ are described in
\ref{sec:nonsup}. 

Thus, assuming that $(\sigma^1,\sigma^2) \to
{\mathbf y}^{(a)}(\sigma)$ (${\mathbf y}^{(a)} \in {\mathbb R}^{(3)}$) is a
parametrization of $\Sigma^{(a)}$, we introduce an auxiliary field
$\xi^{(a)}(\tau,\sigma)$ to exponentiate each functional delta:
\begin{equation}\label{eq:measure}
\delta_{\Sigma^{(a)}}\big[\varphi\big] \;=\; \int \mathcal{D}\xi^{(a)} 
\, e^{i \int d\tau \int d^2\sigma \sqrt{g^{(a)}(\sigma)} \,
\xi^{(a)}(\tau,\sigma) \varphi\big[\tau,{\mathbf y}^{(a)}(\sigma)\big]}
\end{equation}
(no sum over $a$), where $g^{(a)}(\sigma) \equiv \det \big[g^{(a)}_{\alpha\beta}(\sigma)\big]$,
($\alpha,\beta = 1, 2$) is the determinant of the induced metric
$g^{(a)}_{\alpha\beta}$ on the surface, and $\tau \equiv x_0$. 
In terms of the previous parametrization, 
\begin{equation}
g^{(a)}_{\alpha\beta}(\sigma)\;=\; \frac{\partial {\mathbf
y}^{(a)}(\sigma)}{\partial \sigma^\alpha} \cdot \frac{\partial {\mathbf
y}^{(a)}(\sigma)}{\partial \sigma^\beta}\;\;\;({\rm no\; sum\; over} \,a) \;.  
\end{equation}  
Equation (\ref{eq:measure}) guarantees the (necessary)
reparametrization invariance on each surface, assuming that the auxiliary
fields behave as scalars under those transformations. 

Inserting (\ref{eq:measure}) into (\ref{eq:defzsiga1}), we are left with an
equivalent functional integral expression for ${\mathcal
Z}\big[\{\Sigma^{(a)}\}\big]$:
\begin{equation}
{\mathcal Z}\big[\{\Sigma^{(a)}\}\big]\;=\; \int \left( \prod_{a=1}^N
\mathcal{D} \xi^{(a)} \right) \; 
\int \mathcal{D} \varphi \; \exp\Big\{-S_0(\varphi)\,+\,i\, \int
d^4x\,J(x)\,\varphi(x) \Big\}
\end{equation}
where we introduced $J(x) \equiv \sum_{a=1}^N J^{(a)}(x)$, with:
\begin{equation}
J^{(a)}(x)\;=\; \int d\tau d^2 \sigma \; \sqrt{g^{(a)}(\sigma)} \, \xi^{(a)}(\tau,\sigma)
\,\delta(x_0 - \tau) \, \delta^{(3)}[\mathbf{x}-\mathbf{y}^{(a)}(\sigma)]
\;.
\end{equation}

Performing now the (Gaussian) integral over the $\varphi$ field, the result
may be put, in a condensed form, as follows:
\begin{equation}\label{eq:zsigma}
{\mathcal Z}_\Sigma\;=\; {\mathcal Z}_0 \; \times \; \int  \mathcal{D}\xi
\; e^{-S_\Sigma(\xi)}
\end{equation}
where ${\mathcal Z}_\Sigma \equiv {\mathcal Z}\big[\{\Sigma^{(a)}\}\big]$,
and ${\mathcal Z}_0 = \int  \mathcal{D}\varphi \; e^{-S(\varphi_0)}$.

In (\ref{eq:zsigma}), ${\mathcal D}\xi$ denotes the integration measure for
all the auxiliary fields (we assume there is more than one boundary) and
$S_\Sigma$ is a `nonlocal action' for those fields: 
\begin{eqnarray}
S_\Sigma(\xi) &=& \frac{1}{2} \int d\tau  d^2\sigma \int d\tau' d^2\sigma' \;
\sum_{a,b=1}^N \;\Big[  \xi^{(a)}(\tau,\sigma)\,
\nonumber\\  
&\times& {\mathcal M}_{(ab)}(\tau,\sigma;\tau',\sigma')\,
\xi^{(b)}(\tau',\sigma') \Big] \;, 
\end{eqnarray}
where each matrix elements of ${\mathcal M}$ may be expressed in
terms of the scalar field propagator ${\mathcal K}$: 
\begin{equation}
{\mathcal M}_{(ab)}(\tau,\sigma;\tau',\sigma') \;=\; \sqrt{g^{(a)}(\sigma)}
\; {\mathcal
K}(\tau-\tau';\mathbf{y}^{(a)}(\sigma)-\mathbf{y}^{(b)}(\sigma')) \;
\sqrt{g^{(b)}(\sigma')} 
\;,
\end{equation}
which, in $d=3$, may be written as follows:
\begin{eqnarray}
{\mathcal K}(x_0,{\mathbf x}) &=& \int \frac{d \omega}{2\pi} e^{i \omega
x_0 } \, {\widetilde {\mathcal K}}(\omega,{\mathbf x}) \;, \nonumber\\
{\widetilde{\mathcal K}}(\omega,{\mathbf x}) &=&
\int \frac{d^3k}{(2\pi)^3} \, \frac{e^{i {\mathbf k} \cdot {\mathbf
x}}}{\omega^2 + {\mathbf k}^2} \;=\; \frac{ e^{-|\omega| |{\mathbf
x}|}}{4\pi |{\mathbf x}|} \;.
\end{eqnarray}
Taking advantage of the time independence of the physical system
considered, we Fourier transform in time the auxiliary fields, to write
their action in the following way:
\begin{eqnarray}
S_\Sigma &=& \frac{1}{2} \int \frac{d\omega}{2\pi} \,\int d^2\sigma \int d^2\sigma' \;
\sum_{a,b=1}^N \;\Big[  {\tilde\xi}^{(a)*}(\omega,\sigma)\,
\nonumber\\  
&\times& \widetilde{\mathcal M}_{(ab)}(\omega;\sigma,\sigma') \,
\tilde{\xi}^{(b)}(\omega,\sigma') \Big] \;, 
\end{eqnarray}
where the tilde on the fields denotes their corresponding Fourier
transformed versions, and:
\begin{equation}
\widetilde{\mathcal M}_{(ab)}(\omega;\sigma,\sigma') \;=\; \sqrt{g^{(a)}(\sigma)}
{\mathcal K}_{(ab)}(\omega;\sigma,\sigma') \sqrt{g^{(b)}(\sigma')}
\;,
\end{equation}
where
\begin{equation}
{\mathcal K}_{(ab)}(\omega;\sigma,\sigma') \;\equiv\; 
{\widetilde{\mathcal K}}\big[\omega; {\mathbf y}^{(a)}(\sigma) - {\mathbf y}^{(b)}(\sigma')\big]
\;.
\end{equation}
Since the integral over the auxiliary fields is Gaussian, we have:
\begin{equation}\label{eq:zquot}
\frac{{\mathcal Z}_\Sigma}{{\mathcal Z}_0}\;=\; 
\Big\{ \det \big[\widetilde{\mathcal M}_{(ab)}(\omega;\sigma,\sigma')
\delta(\omega-\omega') \big] \Big\}^{-\frac{1}{2}} \;, 
\end{equation}
where the determinant refers to the continuous indices $\omega,\omega'$,
$\sigma,\sigma'$, as well as the discrete ones $a$, $b$.

Then, recalling the relation between the vacuum functional and the vacuum
energy $E_0$ we find for the latter the expression:
\begin{equation}\label{eq:e0quot}
E_0 \;=\; \frac{1}{2} \, \int \frac{d\omega}{2\pi} \, 
{\rm Tr} \log \Big[\widetilde{\mathcal M}_{(ab)}(\omega;\sigma,\sigma') \Big]
\;,
\end{equation}
where the trace affects the $\sigma,\sigma'$ and $a,b$ indices (the trace over 
frequencies has been explicitly dealt with by means of the
integral). We shall reserve the symbol `${\rm tr}$' for the cases where
a trace over just the (continuous) $\sigma,\sigma'$ indices is needed.

Note that no subtraction of the would-be Casimir `self-energies'
has yet been performed; this step will be considered in the next section.
\section{Non superposition}\label{sec:nonsup}
Since we are interested in the Casimir {\em interaction\/} energy, we will
first extract the self-energies of the surfaces. Besides,
those energies are additive quantities, insensitive to the
phenomenon we wish to consider. 

That extraction can be done by factorizing a diagonal matrix constructed from the
$a=b$ elements of $\widetilde{\mathcal M}$:
\begin{equation}
\widetilde{\mathcal M}_{(ab)}(\omega;\sigma,\sigma') \;=\; \int d^2\sigma''
\,\sum_{c=1}^N\, \widetilde{\mathcal D}_{(ac)}(\omega;\sigma,\sigma'') \;
\widetilde{\mathcal T}_{(cb)}(\omega;\sigma'',\sigma')
\end{equation}
where
\begin{equation}
\widetilde{\mathcal D}_{(ab)}(\omega;\sigma,\sigma') \,\equiv\, 
\widetilde{\mathcal M}_{(aa)}(\omega;\sigma,\sigma')
\, \delta_{ab}
\end{equation}
(no sum over $a$). 
By construction, $\widetilde{\mathcal T}(\omega)$ has the matrix elements:
\begin{equation}\label{eq:tab}
\widetilde{\mathcal T}_{(ab)}(\omega;\sigma,\sigma') \;=\; 
\int d^2\sigma'' \,  \widetilde{\mathcal M}^{-1}_{(aa)}
(\omega;\sigma,\sigma'') \; 
\widetilde{\mathcal M}_{(ab)}(\omega;\sigma'',\sigma')
\;,
\end{equation}
(no sum over $a$).
This factorization implies that $\det(\tilde{\mathcal M})\,=\,\det(\tilde{\mathcal D})
\,\det(\tilde{\mathcal T})$; thus, recalling (\ref{eq:zquot}) and
(\ref{eq:e0quot}), we may write:
\begin{equation}\label{eq:selfen}
E_0 \;=\; \sum_{a=1}^N \, E_0^{(aa)} \,+\, E_I \;,
\end{equation}
where 
\begin{equation}
E_0^{(aa)} \;=\; \frac{1}{2} \,  \int \frac{d\omega}{2\pi} \, 
{\rm tr} \log \Big[\widetilde{\mathcal M}_{(aa)}(\omega)\Big]
\;,
\end{equation}
is the Casimir self-energy of the object labelled by the index 
$a$, and:
\begin{equation}
E_I \;=\; \frac{1}{2} \,  \int \frac{d\omega}{2\pi} \, 
{\rm Tr} \log \Big[\widetilde{\mathcal T}(\omega)\Big]
\;.
\end{equation}
As already advanced, the self-energies, as seen from (\ref{eq:selfen}), are
additive. Besides, they do not contribute to the Casimir forces between the
surfaces~\footnote{They do contribute to the Casimir `pressure' on each
surface, though.}, since they are independent of their relative distances. 

Let us then consider the interaction term, $E_I$. It depends on
$\widetilde{\mathcal T} \equiv {\mathcal I} + \widetilde{\mathcal T}'$
where ${\mathcal I}$ is the identity matrix (in both discrete and continuous indices) and
$\widetilde{\mathcal T}'$ has vanishing diagonal ($a=b$) elements. Moreover, for
$a\neq b$, it coincides with $\widetilde{\mathcal T}_{(ab)}$ of
(\ref{eq:tab}). 

As a final step to obtain our main result, we derive a 
different (but equivalent) expression for $\widetilde{\mathcal T}_{(ab)}$,
such that formula for the interaction energy does not contain explicit
factors of the metric.

To that end, we introduce
$G^{(a)}(\omega;\sigma,\sigma')$, the inverse of ${\mathcal
K}_{(aa)}(\omega;\sigma,\sigma')$:
\begin{equation}
\int d^2\sigma''\, {\mathcal K}_{(aa)}(\omega;\sigma,\sigma'')
G^{(a)}(\omega;\sigma'',\sigma') \,=\, \delta^{(2)}(\sigma-\sigma') \;.
\end{equation}  
Then:
\begin{equation}
\widetilde{\mathcal M}^{-1}_{(aa)}(\omega ; \sigma ,\sigma') \;=\;
\frac{1}{\sqrt{g^{(a)}(\sigma)}} \, G^{(a)}(\omega;\sigma,\sigma') \,
\frac{1}{\sqrt{g^{(a)}(\sigma')}} \;, 
\end{equation} 
and:
\begin{equation}\label{eq:tabsimp}
\widetilde{\mathcal T}'_{(ab)}(\omega;\sigma,\sigma') \;=\;
\frac{1}{\sqrt{g^{(a)}(\sigma)}} \, {\mathcal O}_{(ab)}(\omega;\sigma,\sigma') 
\sqrt{g^{(b)}(\sigma')}
\end{equation}
where 
\begin{equation}
{\mathcal O}_{(ab)}(\omega;\sigma,\sigma') \;\equiv\; \left\{
\begin{array}{ccc}
\int d^2\sigma'' \, G^{(a)}(\omega;\sigma,\sigma'') {\mathcal
K}^{(ab)}(\omega;\sigma'', \sigma') &{\rm if} & a\neq b \\
0 &{\rm if}  & a = b
\end{array} \right. \;.
\end{equation}
This way of writing ${\mathcal T}'_{(ab)}$ is rather convenient,
since one can show that the determinants of the metric
cancel, leading to a simpler final expression, depending only on
${\mathcal O}$:
\begin{equation}\label{eq:ei}
E_I \;=\; \frac{1}{2} \,  \int \frac{d\omega}{2\pi} \, 
{\rm Tr} \log \Big[ {\mathcal I} + {\mathcal O}(\omega) \Big]
\;,
\end{equation} 
which we use in our subsequent derivations. This equation is the $N$-body
generalization~\footnote{The existence of this generalization is mentioned
in~\cite{Emig}.} of the so called `$TGTG$' formula for the Casimir
interaction between two bodies applied in~\cite{Emig,Kenet} (see
also~\cite{Balian}), which in our notation reads:
\begin{equation}\label{eq:ei2}
E_I \big( \Sigma^{(1)},\Sigma^{(2)} \big)\,=\, 
 \frac{1}{2} \, \int \frac{d\omega}{2\pi} \, 
{\rm tr} \log \Big[ 1 - G^{(1)}(\omega) {\mathcal K}_{(12)}(\omega)
G^{(2)}(\omega) {\mathcal K}_{(21)}(\omega) \Big]\;.
\end{equation}
The latter is obtained from (\ref{eq:ei}) by expanding in powers of
${\mathcal O}$ and summing up the series for the particular case $N=2$:
\begin{equation}
E_I \big( \Sigma ^{(1)},\Sigma ^{(2)} \big)\;=\; -  \int
\frac{d\omega}{2\pi} \, \sum _{k=1}^\infty \frac{1}{2 k} \,
{\rm tr} \Big\{ \big[ 
{\mathcal O}_{(12)}(\omega) {\mathcal O}_{(21)}(\omega)
\big]^k\Big\}\;,
\end{equation}
what yields (\ref{eq:ei2}). 

Coming back to (\ref{eq:ei}), we note that the form of  $G^{(a)}$ shall
depend, implicitly, on the geometry of each surface, and in general cannot
be evaluated exactly, except in rather simple cases.  However, most
properties we shall deal with in this section are independent of that form.

Expression (\ref{eq:ei}) has immediate analogues in $d\neq3$.
Indeed, in $d=1$, we arrive to a result formally identical to
(\ref{eq:ei}), after one notes that the trace only affects the indices that
label the `surfaces', which in this case are just points labelled by their
coordinates $x^{(a)}$ (no parameters $\sigma$ are involved). Besides, the
kernel $\widetilde{\mathcal K}(\omega;x)$ is now:
\begin{equation}
\widetilde{\mathcal K}(\omega;x) \;=\; \frac{e^{- |\omega| |x|}}{2
|\omega|} \;, 
\end{equation}
and $G^{(a)}$ becomes:
\begin{equation}
G^{(a)}(\omega) = \Big[ \lim_{x,x' \to a} \frac{e^{- |\omega| |x-x'|}}{2
|\omega|} \Big]^{-1} 
 =  2 |\omega| \;,
\end{equation}
independently of $a$.

Finally, in $d=2$ the boundaries are curves $\Gamma^{(a)}$ described by just one
parameter $\sigma$, and:
\begin{equation}
\widetilde{\mathcal K}(\omega;{\mathbf x}) \;\equiv\; \frac{1}{2\pi} \, K_0(
|\omega | |{\mathbf x}|) \;,  
\end{equation} 
where $K_0$ is a modified Bessel function, and $G^{(a)}$ 
is obtained by evaluating the inverse of ${\mathcal K}_{(aa)}$, for which
there is no general expression; we shall however derive its exact form for
a particular case in the next section.

Equipped with (\ref{eq:ei}), we can define a way to `measure' the non-superposition 
effects.  Again, we work in $d=3$, but the results are straightforwardly adapted to
$d\neq 3$:  Assuming that we know $E_I\big(\{\Sigma_{(a)}\}_{a= 1}^N\big)$, the energy
corresponding to $N$ surfaces, we add an extra boundary, $\Sigma_{(N+1)}$,
obtaining a new energy $E_I\big(\{\Sigma_{(a)}\}_{a= 1}^{N+1}\big)$.

If superposition were valid, the difference between the two energies would
be the sum of the interaction energies between $\Sigma_{(N+1)}$ and
$\Sigma_{(a)}$, with $a=1,\ldots,N$. Thus, we introduce:
\begin{equation}
\delta E_I(N) \,\equiv \,E_I\big(\{\Sigma_{(a)}\}_{a= 1}^{N+1}\big) -
E_I\big(\{\Sigma_{(a)}\}_{a= 1}^N\big) 
- \sum_{a=1}^N E_I\big(\{\Sigma_{(N+1)},\Sigma_{(a)}\}\big)\;.
\end{equation} 
Superposition is broken whenever $\delta E_I(N)\neq 0$. Reciprocally, for
the energy of $N$ surfaces to verify superposition we would need: 
$\delta E_I(M) =0$, for $M=2,\ldots,N-1$.

The final ingredient to evaluate $\delta E_I(N)$ is obtained by applying
(\ref{eq:ei}) to the $N+1$ surfaces. Then, we use determinant algebra to 
relate the determinant of the corresponding $(N+1)^{th}$-order matrix to an
$N^{(th)}$-order one:
\begin{equation}
\det \big[{\mathcal I}_{(ab)} + {\mathcal
O}_{(ab)}\big]_{(N+1)\times (N+1)}\;=\;
\det \big[{\mathcal I}_{(ab)} + {\mathcal O}'_{(ab)}\big]_{N\times N}\;,
\end{equation}
where:
\begin{equation}
{\mathcal O}'_{(ab)} \,\equiv\, {\mathcal O}_{(ab)} \,-\,{\mathcal O}_{(a
\; N+1)} {\mathcal O}_{(N+1\; b)} \;.
\end{equation}
Thus, the difference between the energies for $N+1$ and $N$ surfaces may be
put in the form:
\begin{equation}
E_I\big(\{\Sigma_a\}_{a= 1}^{N+1}\big) \,=\,  E_I\big(\{\Sigma_a\}_{a=
1}^N\big) + \frac{1}{2} \int \frac{d\omega}{2\pi} \, 
{\rm Tr} \log \Big[{\mathcal I} - {\mathcal Q} \Big] 
\end{equation}
where
\begin{eqnarray}
{\mathcal Q}_{(ab)}(\omega;\sigma,\sigma') &\equiv&
\int d^2\sigma'' \int d^2\sigma''' \, \sum_{c=1}^N \Big\{ \Big[\big({\mathcal I} +
{\mathcal O}\big)^{-1}\Big]_{(ac)}(\omega;\sigma,\sigma'')
\nonumber\\
&\times& {\mathcal O}_{(c\;N+1)}(\omega;\sigma'',\sigma''') 
{\mathcal O}_{(N+1\; b)}(\omega;\sigma''',\sigma') 
\Big\} \;.
\end{eqnarray}

It is now a matter of algebra to extract the pairs
interaction energy to show that:
\begin{equation}
\delta E_I(N) \;=\; \frac{1}{2}  \int \frac{d\omega}{2\pi} \, 
{\rm Tr} \log[ I + \Lambda(\omega) ] 
\end{equation}
with:
\begin{eqnarray}
\Lambda_{(ab)} & = & \big[I -  {\mathcal O}_{(N+1\;a)} {\mathcal
O}_{(a\;N+1)}\big]^{-1}\;\nonumber\\
&\times& \Big\{ \sum_{c=1}^N [{\mathcal O}\; ( {\mathcal I} + {\mathcal O})^{-1}]_{(ac)}
\;   {\mathcal O}_{(c\;N+1)} {\mathcal O}_{(N+1\;b)} \nonumber\\
&-& {\mathcal O}_{(a\;N+1)}  {\mathcal O}_{(N+1\;b)} + {\mathcal
O}_{(N+1\;a)} {\mathcal O}_{(a\;N+1)} \delta_{ab} \Big\}
\end{eqnarray}
where the $I$ in the first factor is the identity operator on functions
defined in parameter space (while ${\mathcal I}$ also acts on the indices
space), the discrete indices are not summed, and the products are
understood in the operatorial  sense, regarding the kernels as matrix
elements with continuous indices. 

In spite of the fact that the form its rather complicated, we may already
extract some conclusions from it. The most immediate one is
that for the strength of the non-superposition effects to be
small, the magnitude of the matrix elements of ${\mathcal O}$ between
the $(N+1)^{th}$ surface and the previous ones has to be small. 

Moreover, for the correction to be smaller than the superposition terms, 
we also need 
${\mathcal O}_{(a b)}$, for $a,b =1,\ldots N$ to be small, since these
operators also affect the magnitude of those terms. And this is the main conclusion of
this section, namely, that for superposition to be valid, all the matrix
elements of ${\mathcal O}$ have to be small.
We can see, in fact, that when that is the case, the {\em form\/} of the
correction, to lowest order in the matrix elements, does depends on
the matrix elements involving all the boundaries:
\begin{equation}\label{eq:low}
\delta E_I(N) \;\sim \; -\frac{1}{2}  \int \frac{d\omega}{2\pi}
\,\sum_{a,b=1}^N 
{\rm tr}\Big[{\mathcal O}_{(N+1\;a)} {\mathcal O}_{(a b)}
{\mathcal O}_{(b \; N+1)} \Big]\;.
\end{equation}
The smallness of ${\mathcal O}$ is what, on the other hand, renders a
perturbative expansion of the interaction energy possible. 

\section{Perturbative expansion}\label{sec:perturbative}
The condition that the matrix elements of ${\mathcal O}$ are small, is
precisely what one would require in order to expand the interaction energy
in powers of that operator. On the other hand, for ${\mathcal O}$ to be
small, the only assumption available here is that the $N$ surfaces
$\Sigma^{(a)}$ are compact objects, and that the distance between each pair
of surfaces is much bigger than the size of any object.  Under this
assumption, the norm of ${\mathcal O}_{(ab)}$ is much smaller than one,
since the $G^{(a)}$ kernel, is determined by the {\em inverse\/} of
$\widetilde{\mathcal K}$ at small distances, while ${\mathcal K}_{(ab)}$
is, essentially, $\widetilde{\mathcal K}$ at long distances,
and $\widetilde{\mathcal K}$ decreases with the distance. 

Excellent articles exist about the evaluation of the Casimir interaction
energy within the $TGTG$ formula approach, by applying different
expansions{\cite{Milton:2007wz,Emig,Kenet}.  We just present here an
analysis of the perturbative expansion in powers of ${\mathcal O}$, from the point of view of the
non-superposition effects, for the case of $N$ boundaries. 

The expansion yields a series for $E_I$:
\begin{equation}
E_I \;=\; \sum_{l=1}^\infty \, E_{I;l} \;,
\end{equation}
where
\begin{equation}
E_{I;l} \;=\; \frac{(-1)^{l-1}}{2 l} \,  \int \frac{d\omega}{2\pi} \, 
{\rm Tr}  \Big[\big({\mathcal O}(\omega)\big)^l\Big]
\;,
\end{equation}
or:
\begin{equation}
E_{I;l} \;=\; \frac{(-1)^{l-1}}{2 l} \,  \int \frac{d\omega}{2\pi} \, 
\sum_{\scriptscriptstyle{a_1\neq a_2\neq a_3\neq\ldots\neq a_l\neq a_1}} \,
{\rm tr} \Big[
{\mathcal O}_{(a_1a_2)}(\omega)
{\mathcal O}_{(a_2a_3)}(\omega)
\ldots
{\mathcal O}_{(a_la_1)}(\omega)
\Big]
\;.
\end{equation}
This is, essentially, the long distance expansion considered
in~\cite{Milton:2007wz}, although we only deal with the Dirichlet (strong
coupling) case. 

It is worth noting at this point that the absence of explicit factors of
the metric by no means signal a breaking of reparametrization invariance.
Indeed, what happens is that the kernels denoted by $G^{(a)}$ do have a
nontrivial transformation properties under reparametrization, which
compensate for the non invariance of the integrals over the parameters.

Let us study the explicit form of the first few terms in this expansion. The $l=1$
term vanishes, so that the lowest non-trivial order corresponds to $l=2$,
which using (\ref{eq:tabsimp}) becomes:
\begin{eqnarray}
E_{I;2} &=& - \, \frac{1}{4} \,  \int \frac{d\omega}{2\pi} \, 
\sum_{a \neq b} \, \int d^2\sigma \int d^2\sigma'\;
{\mathcal O}_{(ab)}(\omega;\sigma,\sigma')  
{\mathcal O}_{(ba)}(\omega;\sigma',\sigma) \nonumber\\ 
 &\equiv& \sum_{a < b} \, E^{(ab)} 
\end{eqnarray}
where
\begin{eqnarray}\label{eq:quad}
 E^{(ab)} &=& - \frac{1}{2} \int \frac{d\omega}{2\pi} \, 
 \int_{\sigma,\sigma',\sigma'',\sigma'''} G^{(a)}(\omega;\sigma,\sigma') \;
{\mathcal K}^{(ab)}(\omega;\sigma',\sigma'') \nonumber\\
&\times&
G^{(b)}(\omega;\sigma'',\sigma''')\; {\mathcal
K}^{(ba)}(\omega;\sigma''',\sigma) \;.
\end{eqnarray}
To this order, the total energy is obtained as the sum of
`interaction energies' corresponding to the pairs, in a sort of
`superposition principle'.

This property is violated in the next order term:
\begin{eqnarray}
E_{I;3} &=& \frac{1}{6} \,  \int \frac{d\omega}{2\pi} \, 
\sum_{a, b, c} \, \int d^2\sigma \int d^2\sigma' \int d^2\sigma''
\nonumber\\
&\times&
{\mathcal O}_{(ab)}(\omega;\sigma,\sigma')  
{\mathcal O}_{(bc)}(\omega;\sigma',\sigma'')  
{\mathcal O}_{(ca)}(\omega;\sigma'',\sigma) \nonumber\\ 
 &\equiv& \sum_{a < b < c} \, E^{(abc)} 
\end{eqnarray}
where we have introduced a `three-body energy interaction', $E^{(abc)}$:
\begin{equation}
 E^{(abc)} \;=\; \int \frac{d\omega}{2\pi} \; {\rm tr} \Big[
 G^{(a)}(\omega){\mathcal K}^{(ab)}(\omega)
G^{(b)}(\omega){\mathcal K}^{(bc)}(\omega) 
G^{(c)}(\omega){\mathcal K}^{(ca)}(\omega) \Big] \;.
\end{equation}
Incidentally, this correction coincides with (\ref{eq:low}) when one
considers $N+1$ surfaces, as it should be, since on should expect that the
lowest order violation to the non-superposition comes from the lowest
non-quadratic term in the energy.

A fundamental ingredient in the calculation of the different terms in the
expansion for $E_I$ is the kernel $G^{(a)}(\omega;\sigma,\sigma')$. The
form of that kernel depends strongly on the number of spatial dimensions
as well as on the shape of the surface itself. Universal statements can
only be made if more assumptions about the surfaces are made.
However, based on the same assumption used to perform the series expansion, we
may simplify the previous expressions further. Indeed, denoting by ${\mathbf
x}^{(a)}$ the barycenter of the $\Sigma^{(a)}$ surface, we can, in
the expressions above, use the approximation:
\begin{equation}
 {\mathcal K}^{(ab)}(\omega;\sigma',\sigma'') \simeq 
{\widetilde{\mathcal K}}(\omega; {\mathbf x}^{(a)} - {\mathbf x}^{(b)} ) \;.
\end{equation}  
This is justified by the following reason: we are assuming that $|{\mathbf
x}^{(a)} - {\mathbf x}^{(b)}| >> R^{(a)}, R^{(b)}$, where $R^{(a)}$ denotes
the minimum radius for a sphere $S^{(a)}$, centered at ${\mathbf x}^{(a)}$,
which encloses $\Sigma_a$. Then we may replace ${\mathbf
y}^{(a)}(\sigma) \to {\mathbf x}^{(a)}$ and ${\mathbf y}^{(b)}(\sigma') \to
{\mathbf x}^{(b)}$, since ${\mathcal K}^{(ab)}(\omega;\sigma',\sigma'')$ is
(under the previous assumptions) approximately constant inside $S^{(a)}$. 

Using this approximation inside the expression for $E^{(ab)}$, we see that
it may be written as follows:
\begin{equation}
 E^{(ab)} \;\simeq\;  \int \frac{d\omega}{2\pi} \,\int d^3x
\int d^3y \; 
\rho^{(a)}(\omega;{\mathbf x}) V(\omega;{\mathbf x}- {\mathbf y})
\rho^{(b)}({\mathbf y}) \;,
\end{equation}
where we introduced:
\begin{eqnarray}
\rho^{(a)}(\omega;{\mathbf x}) &\equiv & 
q_a (\omega) \; \delta^{(3)}({\mathbf x}-{\mathbf x}^{(a)}) \nonumber\\
q_a(\omega) &\equiv &  \int d^2\sigma \int d^2\sigma' \; G^{(a)}(\omega;\sigma,\sigma')
\end{eqnarray}
and
\begin{equation}
V(\omega;{\mathbf x}- {\mathbf y})
\;\equiv\; -\frac{1}{2} \;\Big[ {\widetilde K}(\omega;{\mathbf x}- {\mathbf
y})\Big]^2 \;.
\end{equation}
Thus, at this order, we see that the interaction energy for the $a,\,b$
pair, may be regarded as arising from integral over $\omega$ of the
interaction energy for a set of pointlike charges located at ${\mathbf
x}^{(a)}$ and ${\mathbf x}^{(b)}$, whose strengths $q_a(\omega)$ and
$q_b(\omega)$ are determined by the geometry of the respective surface.  

On the other hand, the explicit form of the interaction potential is:
\begin{equation}
V(\omega;{\mathbf x}- {\mathbf y})
\;\equiv\; -\frac{1}{2} \, \frac{e^{-2 |\omega| |{\mathbf x} - {\mathbf
y}|}}{ (4 \pi)^2 |{\mathbf x} - {\mathbf y}|^2} \;,
\end{equation}
hence, the interaction is always attractive.  The integrals over ${\mathbf
x}$ and ${\mathbf y}$ have been used in order to make it clear that each
surface behaves as a sort of point-like charge. Of course, the same
approximation may be used to simplify the form of the higher order terms.

The form of $G^{(a)}$ is not known exactly in general, except for particular
situations, like the $d=1$ case, which we consider now:
\subsection{$d=1$}
As a first test, we consider the case of two mirrors in $1+1$ dimensions.
The operator ${\mathcal O}$ is just an $\omega$-depending matrix, with
matrix elements ${\mathcal O}_{(ab)}(\omega) = e^{-|\omega|
|x^{(a)}-x^{(b)}|}$; the exponential decay assures the convergence of the
perturbative expansion, regardless of the relative distances between the
mirrors.
 
In this situation, the first (superposition) expression for the energy
corresponding to two point-like objects (mirrors) located at $x^{(1)}$ and
$x^{(2)}$ adopts the form:
\begin{equation}
 E^{(12)} \;=\;  \int \frac{d\omega}{2\pi} \, 
(2 |\omega| )^2 V(\omega; x^{(1)}- x^{(2)})  \;,
\end{equation}
where 
\begin{equation}
V(\omega; x^{(1)}- x^{(2)}) \,=\,-\frac{1}{2} \; \frac{e^{-
2|\omega||x^{(1)}- x^{(2)}|}}{ ( 2 |\omega|)^2}
\end{equation}
Assuming that the distance between the mirrors is $a$, we see that.
\begin{equation}
E^{(12)} \;=\; -\frac{1}{2} \, \int_{-\infty}^{+\infty}
\frac{d\omega}{2\pi} \, e^{- 2|\omega| a} \,=\; -\frac{1}{4 \pi a} \;=\; -
\frac{0.07958}{a} \, \;,
\end{equation}
to be compared with the exact result: $E = - \frac{\pi}{24 a} \simeq -
\frac{0.1309}{a}$, which is bigger by approximately a sixty percent. 

It is possible to calculate, for this case, all the higher order
corrections exactly; only the even orders yield non-vanishing
contributions, which are given by:
\begin{equation}
E_{I;2l}\;=\; - \frac{1}{2l} \, \int_{-\infty}^{+\infty}
\frac{d\omega}{2\pi} \, \big(e^{- 2|\omega| a}\big)^{2 l} \,=\; -\frac{1}{4
l^2 \pi a} \;.
\end{equation} 
Then one sees that their sum:
\begin{equation}
\sum_{l=1}^\infty E_{I;2l}\;=\; - \frac{1}{4 \pi a} \,
\sum_{l=1}^\infty \frac{1}{l^2} \,=\,  - \frac{1}{4 \pi a} \frac{\pi^2}{6} \,=\, 
- \frac{\pi}{24 a} \;,
\end{equation} 
which is the exact result. 

Besides, when more than two mirrors are considered, the energy becomes
equal to the sum of the Casimir energies corresponding to the pairs formed
by neighboring mirrors:
\begin{equation}
E_I = -\frac{\pi}{24} \, \sum_{a=1}^{N-1} \frac{1}{|x^{(a+1)} - x^{(a)}|}
\;.
\end{equation} 
\subsection{$d=2$}
 For a radius $R$ circle, using the angle $\phi$ as parameter, we
find:
\begin{equation}
G^{(a)}(\omega;\phi,\phi')\;=\; \frac{1}{2\pi}
\,\sum_{n=-\infty}^{+\infty} \, \frac{e^{i n (\phi -
\phi')}}{I_{|n|}(|\omega| R)  K_{|n|}(|\omega| R)}\;.
\end{equation}
For an infinite line, parametrized by $\sigma \in (-\infty,+\infty)$, the
result is instead:
\begin{equation}
G^{(a)}(\omega;\sigma,\sigma')\;=\; 2 \, \big(- \frac{\partial^2}{\partial
\sigma^2} + \omega^2\big) K_0(|\omega| |\sigma - \sigma'|) \;.
\end{equation}

It is straightforward to check that, in both cases, one is in a situation
of an ${\mathcal O}$ with small norm.
Thus, in the perturbative expansion, when the surfaces are bounded
and very far away, using the approximation that follows from:  
\begin{equation}
{\mathcal K}_{(ab)}(\omega;|{\mathbf x}^{(a)} - {\mathbf x}^{(b)} |)\sim 
\widetilde{\mathcal K}(\omega;|{\mathbf x}^{(a)} - {\mathbf x}^{(b)} |) 
\end{equation}
we may obtain an approximate expression for the case
of $N$ circles. Denoting by $\eta^{(a)}$ and ${\mathbf x}^{(a)}$ the center
and radius of each circle, and assuming that $|{\mathbf x}^{(a)}-{\mathbf
x}^{(b)}| >> {\rm max}\{\eta^{(a)}\}$, we have for the pair interaction energy:
\begin{equation}
E^{(ab)} \;\sim\; - \frac{1}{2} \, \int \frac{d\omega}{2\pi} \, 
\frac{[K_0(|\omega||{\mathbf x}^{(a)}-{\mathbf x}^{(b)}|)]^2}{I_0(|\omega|\eta^{(a)}) 
K_0(|\omega|\eta^{(a)}) I_0(|\omega|\eta^{(b)}) K_0(|\omega|\eta^{(b)})} \;.
\end{equation} 
For the case of just
two circles, $1$ and $2$, say, one can sum the series corresponding to the different powers
of ${\mathcal O}$. The result is:
\begin{equation}
E_I \;\sim\; - \frac{1}{2} \, \int \frac{d\omega}{2\pi} \, 
\log \Big[1 - \frac{[K_0(|\omega||{\mathbf x}^{(a)}-{\mathbf x}^{(b)}|)]^2}{I_0(|\omega|\eta^{(a)}) K_0(|\omega|\eta^{(a)}) 
I_0(|\omega|\eta^{(b)}) K_0(|\omega|\eta^{(b)})} \Big]\;.
\end{equation} 

\subsection{$d=3$}
We now deal with the case of surfaces in $d=3$.  
The long distance approximation requires the evaluation of the integral of
$G^{(a)}$ over the parameters; for a case of a sphere, that integral is:
\begin{equation}
q(\omega) = \frac{4 |\omega| R^2}{I_{1/2} (|\omega| R) K_{1/2} (|\omega| R)
} \;.
\end{equation}
This is smaller than ${\mathcal K}_{(ab)}$ for distant surfaces, as a
straightforward test shows.

If one assumes instead that the surfaces are really composed of
small, weakly coupled surface elements, we may in fact use local
approximations for the $G^{(a)}$ kernels.  In this case, a local
approximation means that the kernel is concentrated around $\sigma =
\sigma'$: \begin{equation}
G^{(a)}(\omega;\sigma,\sigma')\;\sim \; \eta^{(a)}(\omega, \sigma) \;
\delta^{(2)} (\sigma-\sigma') \;.
\end{equation}
where $a =1,2$, and $\eta^{(a)}$ will be determined now:
Recalling that $G^{(a)}$ is defined as the inverse of ${\mathcal
K}_{(aa)}(\omega;\sigma,\sigma')$, we explore the form the latter in the
neighborhood of a given point in the surface $\Sigma_a$, the one
characterized by the parameter $\sigma$: ${\mathbf y}^{(a)}(\sigma)$. Close
to that point, we derive the approximate expression:
\begin{eqnarray}
{\mathcal K}_{(aa)}(\omega;\sigma,\sigma') &\sim& \int \frac{d^2k_\parallel}{(2\pi)^2} \, \frac{e^{i {\mathbf k}_\parallel
\cdot \partial_\alpha {\mathbf y}^{(a)}(\sigma)
(\sigma^\alpha-\sigma'^\alpha)}}{ 2 \sqrt{{\mathbf k}_\parallel^2 +
\omega^2}} \nonumber\\
&\sim& \frac{1}{4\pi \sqrt{g_{\alpha\beta}^{(a)}(\sigma)
\delta\sigma^\alpha \delta\sigma^\beta}} \, \exp\big[- |\omega|
\sqrt{g_{\alpha\beta}^{(a)}(\sigma)
\delta\sigma^\alpha \delta\sigma^\beta}\big] \;,
\end{eqnarray}
where ${\mathbf k}_\parallel$ is the projection of the momentum along the
tangent plane at the point ${\mathbf y}^{(a)}(\sigma)$, and
$\delta\sigma^\alpha \equiv \sigma^\alpha - \sigma'^\alpha$. 

In the assumption that there is no appreciable momentum flux between the
different surface elements, we end up with the expression:
\begin{equation}
{\mathcal K}_{(aa)}(\omega;\sigma,\sigma') \;\sim \; \frac{1}{2
\sqrt{g^{(a)}(\sigma)} |\omega|} \,
\delta^{(2)}(\sigma-\sigma') \;.
\end{equation}
This yields:
\begin{equation}
G^{(a)}(\omega;\sigma,\sigma')\;\sim\; 2 \sqrt{g^{(a)}(\sigma)} |\omega| \, \delta^{(2)}(\sigma-\sigma') 
\;\;\Rightarrow \; \eta(\omega,\sigma) = 2 |\omega| \,g^{(a)}(\sigma)\;.  
\end{equation}

Let us first assume that we have two surfaces, $\Sigma^{(1)}$ and $\Sigma^{(2)}$,
and consider the second order expression for the interaction energy, using
the local approximation for the kernels $G^{(1)}$ and $G^{(2)}$.
We see that their interaction energy at this order becomes:
\begin{equation}
E^{(12)} \;=\; - \frac{2^2}{2} \int \frac{d\omega}{2\pi} \,
\omega^2 \, \int d^2\sigma  \sqrt{g^{(1)}(\sigma)} \int d^2\sigma'
\sqrt{g^{(2)}(\sigma')} \big[{\mathcal
K}^{(12)}(\omega;\sigma,\sigma')\big]^2
\end{equation}
which, performing the integration over $\omega$, results in the following
expression:
\begin{equation}
E^{(12)} \;=\; \int d^2\sigma  \sqrt{g^{(1)}(\sigma)} \int d^2\sigma'
\sqrt{g^{(2)}(\sigma')} V(\sigma,\sigma')
\end{equation}
where:
\begin{equation}
V(\sigma,\sigma') \;=\; - \frac{1}{32 \pi^3} \, \frac{1}{|{\mathbf
y}^{(1)}(\sigma) - {\mathbf y}^{(2)}(\sigma')|^5}
\end{equation}
which looks like a kind of local-potential, van
der Waals like interaction.

Finally, let us consider the case of infinite parallel plates, within the
quadratic approximation, using two different approaches. 
Obviously, in this case, the planes cannot be regarded as small surfaces
and, even though the superposition approximation may be valid, certainly
the planes cannot be regarded as point-like objects. 

It is clear that, using as parameters the coordinates ${\mathbf
x}^{(a)}_\parallel$ on each mirror, we have:
\begin{equation}
G^{(a)}(\omega;{\mathbf x}^{(a)}_\parallel,{\mathbf y}^{(a)}_\parallel)\;=\; 
\int \frac{d^2k_\parallel}{(2\pi)^2} \; 2 \, \sqrt{k_\parallel^2 +
\omega^2} \; e^{i {\mathbf k}_\parallel \cdot ({\mathbf x}^{(a)}_\parallel -{\mathbf y}^{(a)}_\parallel)} \;,
\end{equation}
while for ${\mathcal O}$ the result is:
\begin{equation}
{\mathcal O}_{(ab)}(\omega;{\mathbf x}^{(a)}_\parallel,{\mathbf y}^{(b)}_\parallel)\;=\; 
\int \frac{d^2k_\parallel}{(2\pi)^2} \; e^{-\sqrt{k_\parallel^2 + \omega^2}
|z^{(a)} - z^{(b)}|+ i {\mathbf k}_\parallel \cdot ({\mathbf x}^{(a)}_\parallel -
{\mathbf y}^{(a)}_\parallel)} \;,
\end{equation}
where $z^{(a)}$ is the position (on the third axis) of each plane. We see
that, even in this case, the norm of the operator is small. 

Indeed, inserting this into (\ref{eq:quad}), we get for ${\mathcal E}_0$, the
energy per unit area:
\begin{equation}
{\mathcal E}_0 \;=\; - \frac{1}{2}  \int \frac{d\omega}{2\pi} \,
\int \frac{d^2k_\parallel}{(2\pi)^2} \;e^{- 2 a \sqrt{{\mathbf
k}_\parallel^2 + \omega^2}} \;=\; \frac{1}{16 \pi^2 a^3} \;\simeq\; -0.00633 
\, a^{-3} \;,
\end{equation}
to be compared with the exact result, that is ${\mathcal E}_{I;2} \simeq
-0.006854 \, a^{-3}$, what is a signal that the corrections are small.

If, on the other hand, we imagine each mirror as composed of
weakly interacting infinitesimal surface elements (not a conductor), and 
apply the superposition result to a system composed of all the surface
elements, then the energy per unit area to the first non-trivial order, ${\mathcal
E}_{I;2}$, may be obtained by integrating the interaction energy between a
single point on a mirror and all the points in the other. This corresponds
to the following integral:
\begin{equation}
{\mathcal E}_{I;2}\;=\;  4 \, \int \frac{d\omega}{2\pi}\,  \omega^2 
\int d^2x_\parallel \, V(\omega; \sqrt{a^2 + {\mathbf x}_\parallel^2}) \;.
\end{equation}

The integral can be evaluated exactly, yielding:
\begin{equation}
{\mathcal E}_{I;2}\;=\; -\frac{1}{24 \pi^2 a^3} \;\simeq\; - 0.00422 \, a^{-3}
\end{equation}
which is different than the previously obtained result, as it corresponds
to a different material.

\section{Conclusions}\label{sec:concl}
We have obtained an expression that measures the departure from
superposition in the interaction Casimir energy corresponding due to a
massless scalar field in the presence of $N > 2$ Dirichlet surfaces.  We
have found that the most general condition under which the
non superposition effects can be regarded as small corresponds to a number of small
surfaces separated by long distances. 
Under this assumption, one may construct a perturbative expansion, as a
series in the operator ${\mathcal O}$.

The condition on that operator manifests itself in a
different fashion, depending on the number of spatial dimensions. In $d=1$,
since the size of the mirrors is zero, one is in the best possible
situation, namely, the perturbative expansion is always reliable.

In $d=2$ and $d=3$, on the other hand, one can always obtain conditions
under which the expansion should be reliable (although the rate of convergence depends
on $d$). 

An interesting conclusion one can extract is that the knowledge of the
interaction energy for $N$ surfaces is useful to calculate the one for
$N+1$ surfaces only when all the surfaces are widely separated.
\section*{Acknowledgements}
C.C.T, C.D.F. and E.L.L. thank CONICET, ANPCyT and UNCuyo for financial support. 
\newpage


\begin{thebibliography}{bib}
\bibitem{rev}
G. Plunien, B. M\"uller, and W. Greiner, Phys. Rep. \textbf{134},
87 (1986); V.\ M.\ Mostepanenko and N. N. Trunov, {\it The
Casimir Effect and its Applications} (Clarendon, London, 1997); M.
Bordag, {\it The Casimir Effect 50 Years Later} (World Scientific, Singapore, 1999);
M. Bordag, U. Mohideen, and V. M. Mostepanenko, Phys. Rep.
\textbf{353}, 1 (2001); K. A. Milton, {\it The Casimir Effect:
Physical Manifestations of the Zero-Point Energy} (World
Scientific, Singapore, 2001); S. Reynaud {\it et al.}, C. R. Acad.
Sci. Paris \textbf{IV-2}, 1287 (2001); K. A. Milton, J. Phys. A:
Math. Gen. \textbf{37}, R209 (2004); S.K. Lamoreaux, Rep. Prog.
Phys. \textbf{68}, 201 (2005); Special Issue {\it "Focus on Casimir Forces"},
New J. Phys. \textbf {8} (2006).
\bibitem{milonni}P.\ Milonni, {\it The Quantum Vacuum} (Academic Press,
San Diego, 1994);
\bibitem{kardar} H.\ Li and M.\ Kardar, Phys.\ Rev.\ {\bf A46}, 6490
(1992).
\bibitem{kardar2} T.\ Emig, A.\ Hanke, R.\ Golestanian, and M.\ Kardar,
Phys.\ Rev.\ Lett.\ {\bf 87}, 260402 (2001); ibidem Phys.\ Rev.\
{\bf A67}, 022114 (2003)
\bibitem{recent}T.\ Emig, R.\ L.\ Jaffe, M.\ Kardar and  A.\ Scardicchio,
Phys.\ Rev.\ Lett.\ {\bf 96}, 080403 (2006); M.\ Bordag,
hep-th/0602295.
\bibitem{zinn}J.\ Zinn-Justin, {\em Quantum Field Theory and Critical Phenomena}, Oxford Science
Publications, 4th. Ed., (2002).
\bibitem{Emig}T.\ Emig, N.\ Graham, R.\ L.\ Jaffe, and M.\ Kardar, Phys.\
Rev.\ Lett.\ {\bf 99}, 170403 (2007).
\bibitem{Kenet}O.\ Kenneth and I.\ Klich, Phys.\ Rev.\ B{\bf 78}, 014103 (2008).
\bibitem{Balian}
R.~Balian and B.~Duplantier,
Annals Phys.\  {\bf 112}, 165 (1978).
\bibitem{Milton:2007wz}
K.~A.~Milton and J.~Wagner,
J.\ Phys.\ A  {\bf 41}, 155402 (2008).
\end{thebibliography}
\end{document}